# Exceptional high Seebeck Coefficient and Gas-Flow-Induced Voltage in Multilayer Graphene

Xuemei Li, Jun Yin, Jianxin Zhou, Qin Wang, Wanlin Guo[†]

State Key Laboratory of Mechanics and Control of Mechanical Structures and the Key Laboratory for Intelligent Nano Materials and Devices of the Ministry of Education, Institute of Nanoscience, Nanjing University of Aeronautics and Astronautics, Nanjing 210016, China

**TABLE OF CONTENTS GRAPHIC**

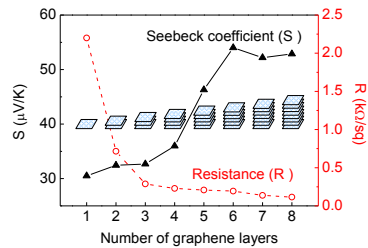

**Abstract:** Monolayer graphene shows Seebeck coefficient several times and gas-flow-induced voltage twenty times higher than that of bulk graphite. Here we find that the Seebeck coefficient of multilayer graphene increases monotonically with increasing layer and reaches its peak value at hexa-layer ~77% higher than for monolayer and then decreases, although the electric resistance decreases monotonically with increasing layer. The flow-induced voltage is significantly higher in 2, 4, 5, 6, 7 layered graphene than in 1, 3, 8 layered one, against the prevailing view that it should be proportional to Seebeck coefficient. These thickness effects are also in sharp contrast to that in continuous aluminum nanofilms.

---

[†] Correspondence should be addressed to W. G. (wlguo@nuaa.edu.cn)

Graphene, a distinct two-dimensional carbon atomic layer, has attracted enormous research interest due to its large specific surface area, high intrinsic mobility,[1,2] structural transparency[3] and mechanical strength[4] and shows great potentials in information and energy technologies.[5,6] Monolayer graphene can be split off from bulk graphite by mechanical or chemical cleavage methods,[2,7] directly grown in large area on surfaces of many materials,[8-10] and has a thermal conductivity above 2000 $Wm^{-1}K^{-1}$ significantly higher than other carbon allotropes.[11] The Seebeck coefficient, specifying the ability of an induced thermoelectric voltage in response to a temperature difference across the material, can be used to generate electricity, measure temperature or control the temperature of objects.[12] Especially interesting, the interplay of the Seebeck effect and Bernoulli's principle can lead to gas-flow-induced voltage.[13] We previously found that the gas-flow-induced voltage in monolayer graphene is twenty-folds over that in bulk graphite, although the Seebeck coefficient of monolayer graphene is about 6 times of that of bulk graphite.[14]

Recently, unique properties, including thermal properties of multilayer graphene have been revealed.[11,15,16] The structure of the low-energy electronic dispersion is linear for monolayer graphene,[17] while in bilayer graphene, the shape of the energy bands is quadratic[18] and becomes cubic in ABC-stacked trilayer.[19] In multilayer graphene, stacking order also provides an important freedom for tuning its electronic properties.[20, 21] However, how the Seebeck coefficient and gas-flow-induced voltage would change with increasing layer from mono- to multi-layered graphene remains completely elusive. Here we find strong and abnormal layer-dependent Seebeck effect and ability for gas-flow-induced voltage in mono- to octa-layered graphene. The Seebeck coefficient does not decrease directly with increasing layer to that of graphite, but increases monotonically with increasing layer and reaches its peak value in hexalayer graphene which is much higher than that of bulk graphite. What is more, the gas-flow-induced voltage in mono- to octa-layered graphene is not in proportion to the Seebeck coefficient, in sharp contrast with the prevailing view proposed by Sood *et.al*.[13] The results shed new light on the unique layer-dependent properties of multilayer materials.

**RESULTS AND DISCUSSION**

In this work, monolayer graphene samples were synthesized by chemical vapor deposition (CVD) method on copper foils (25 μm, Alfa Aesar, item No.13382) using methane as the precursor.[10] After CVD growth, the as-grown graphene samples were transferred onto a polyester terephthalate (PET) substrate by using polymethyl methacrylate (PMMA) as mediator.[22] The quality of the graphene samples was characterized by Raman spectrum at room temperature with a Renishaw spectrometer equipped with a microscope of 50 times magnification and a solid-state laser ( λ=514 nm ) at an intensity of 1 mW. The Raman spectrum of graphene in Fig. S1 shows two large peaks, the G band at ∼1587 $cm^{-1}$ and the 2D band at ∼2686 $cm^{-1}$. The ratio of the G to 2D peak intensity ($I(G)/I(2D)$) is less than 0.3, and the full width at half-maximum (fwhm) of the 2D band is ∼32 $cm^{-1}$, showing the graphene as monolayer. The defect induced D band is nearly undetectable in our sample, indicating a low concentration of defects and high graphene quality.[23] Multilayer graphene stacks from bi- to octa- layers were obtained by superposing the monolayer graphene with random orientation using layer-by-layer transfer-printing techniques. Two terminal electrodes of copper foil were then adhered by silver emulsion for each of the graphene sheet. The exposed part of the graphene sample was adjusted to 15 mm in length and 8 mm in width between the two end electrodes. The current-voltage (*I-V*) curves for all the graphene samples with different layers are linear as shown in Fig. S2, indicating perfectly ohmic contact and the metal characteristics of the samples. The voltage signal is recorded in real time using a KEILTHELY 2010 multimeter, which is controlled by a Labview-based data acquisition system.

First, we measured the Seebeck coefficient of the graphene samples. Without specific statements, the mentioned Seebeck coefficient in the following discussions is referenced to the copper electrode. As shown by the inset of Figure 1a, the left terminal of the sample was electrically heated up to a specific temperature, then the electric power was turned off and the sample was naturally cooled down gradually in the ambient environment. During the cooling process, the temperature at the two end electrodes of the samples and the voltage between them were recorded simultaneously. The heating induced voltage is shown in Figure 1a for the mono- to tetra-layered graphene samples and Figure 1b for penta- to octa-layered graphene samples as a function of the temperature difference between the end electrodes ($\Delta T = T_L - T_R$). The date shows excellent linear relationship, from which the Seebeck coefficient can be easily evaluated. Figure 1c shows the obtained Seebeck coefficient and the square resistance of the graphene samples against the layer number. The graphene resistance decreases monotonically from 2.2 kΩ/□ for monolayer graphene to 115 Ω/□ for octa-layered graphene. The Seebeck coefficient increases monotonically from 30.5 µV/K for monolayer graphene to 56 µV/K for hexa-layered graphene, which is ~77% higher than the monolayer; but then decreases slightly for hepta- and octa-layered graphene. The deduced positive Seebeck coefficient of our monolayer sample is consistent with the theoretical prediction and the fact that graphene can be easily p-doped in the ambient environment for the adsorption of oxygen and water molecules.[24,25] This layer-dependence of Seebeck coefficient is peculiarly and unexpected. The Seebeck coefficient of graphene is determined by the interplay of the band structure, the defects or disorder induced scattering, the temperature and the Landau quantization.[16] In bilayer graphene, the tunable bandgap and potential-energy differences between two layers can enhance the Seebeck effect,[26] but these factors cannot explain the layer-dependent Seebeck effect in the multilayer graphene.

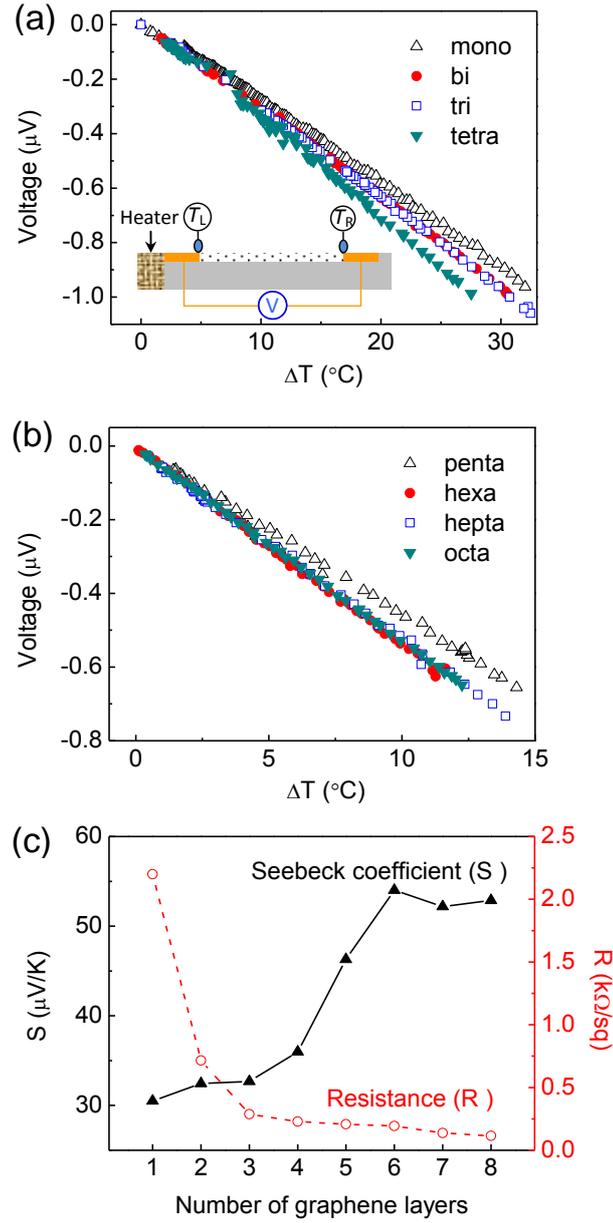

**Figure 1**. Seebck effect in multilayer graphene. Linear variation of the Seebeck voltage with temperature difference along the (a) mono- to tetra-layered and (b) penta- to octa-layered graphene samples. The inset in (a) shows the setup for measurement of the Seebeck effect and electric resistance. (c) Layer-dependent Seebeck coefficient and resistances of multilayer graphene.

We further examined the relationship between gas-flow-induced temperature distribution and the induced voltage in monolayer graphene sample. Figure 2a shows the temperature distribution along the nitrogen flow direction at 68 m/s. The inset shows a schematic layout of the experimental setup. The local temperature was measured by a thermocouple spaced close to the sample surface. The local temperature decrease monotonously from 22.1 ℃ to 19.6 ℃ along the gas flow, in a distance about 15 mm. Then it rises gradually due to the influence of room temperature (~25 ℃). It shows how the temperature changes along the gas flow, which should be caused by the pressure gradient according to Bernoulli's principle. In Figure 2b, the relationship between the gas-flow-induced voltage and the gas-flow-induced temperature difference was compared with the curve of the heating induced voltage against heating induced temperature difference. The good coincidence between them indicates that the gas-flow-induced voltage in monolayer graphene is mainly attributed to the temperature difference caused by the gas flow, confirming the role of the interplay of Seebeck effect and Bernoulli's principle.

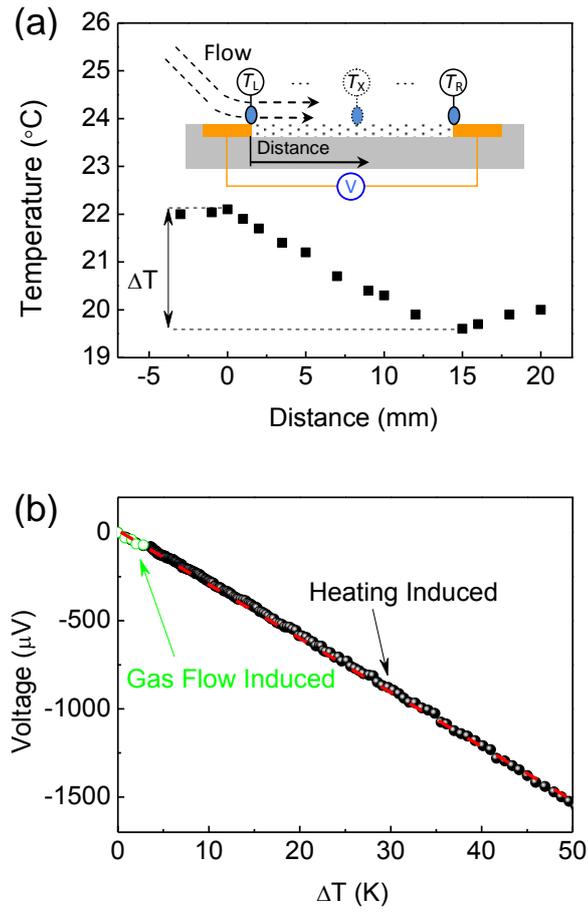

**Figure 2**. (a) Nitrogen-flow-induced temperature distribution along the gas flow direction at the velocity of 68 m/s. The inset shows the setup for the measurement of temperature distribution. (b) Flow (green blank circles) and heating (black dots) induced voltage against the corresponding temperature difference between the two ends of the graphene sample ($\Delta T = T_L - T_R$). Linear fitting (red dash line) of the data yields a Seebeck coefficient $S = 30.5$ μV/K relative to copper.

We then conducted measurement on the gas-flow-induced voltage in multilayer graphene at various nitrogen gas flow velocities. Figure 3a shows the dependence of the induced voltage on the flow velocity. Similar to monolayer graphene, for all the multilayer graphene samples, the gas-flow-induced voltage increases monotonically with increasing flow velocity. However, the gas-flow-induced voltage does not change in the way with the number of graphene layers as the electric resistance or the Seebeck coefficient as shown in Fig. 1c. Instead, as shown clearly by the inset of Fig. 3a, the gas-flow-induced voltage in mono-, tri- and octa-layered graphene is distinguishingly lower in comparing to that in bi-, tetra-, penta-, hexa- and hepta-layered graphene at high flow velocity (156 m/s in the inset). The relationship of the induced voltage in the multilayered graphene samples at flow velocities of 71, 82, 101 and 193 m/s is also shown in Figures S3a-S3d, respectively.

The dependence of the induced voltage on square of the Mach number ($M^2$) is shown in Figure 3b. Similar to the finding by Sood *et.al*.[13] and the trend in monolayer graphene as reported in our previous work,[14] the gas-flow-induced voltage is proportional to the square of the Mach number, where $M = v/c$, $v$ is the gas flow velocity, and $c$ is the sound velocity in medium (353m/s for nitrogen at 300 K), and separating into two linear regions with different slopes. However, the slope in the high velocity region with $M^2 > 0.06$ is larger than that in the low velocity region with $M^2 < 0.06$ for all the samples with bi- to octa-layer, in sharp contrast with the trend reported previously that the slope in the high velocity region is lower.[13,14]

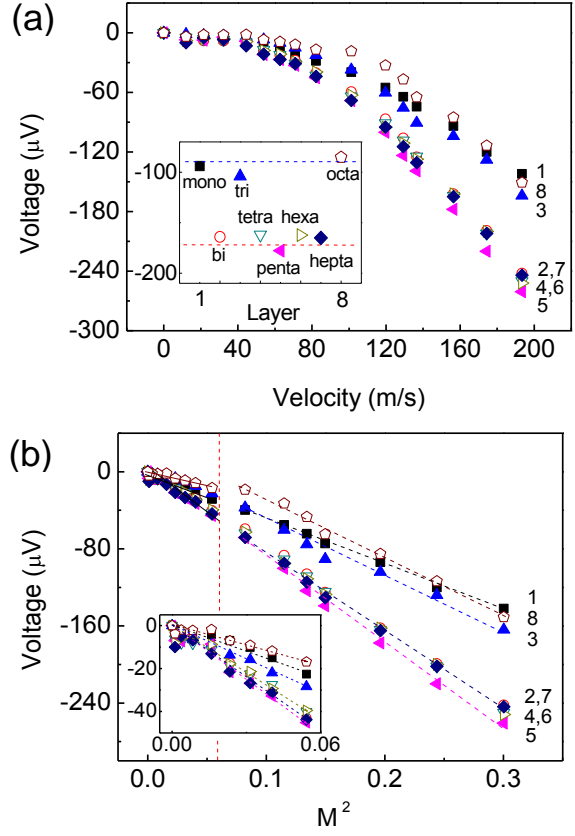

**Figure 3.** Variation of nitrogen gas-flow-induced voltage in the multilayer graphene samples with (a) flow velocity, and (b) square of March number $M^2$. The inset in (a) is the induced voltage at velocity of 156 m/s. The inset in (b) shows the expanded plot of low velocity region.

The ability to produce voltage in gas flow can be expressed more clearly by the ratio of the gas-flow-induced voltage $V_{GF}$ over $M^2$, as shown in Figures 4a and 4b for all the few-layered samples in the low and high velocity regions, respectively. According to the prevailing view proposed by Sood *et al.*,[13] the temperature difference caused by adiabatic flow of a gas can be approximatively deduced as $\Delta T \propto \frac{1}{2}T_0\gamma M^2$ at low velocity, and $\Delta T \propto \frac{1}{2}T_0(\gamma-1)M^2$ at high velocity, where $T_0$ is the environmental temperature and $\gamma$ is the heat capacity of the gas, with $\gamma = 1.404$ for the nitrogen gas here. Thus, the ratio $V_{GF}/M^2$ in the low velocity region should be significantly larger than that in the high velocity region. This has been proven for monolayer graphene in flow of different gases.[14] Here we also find in the monolayer sample that the ratio of $V_{GF}/M^2 = 475$ in the high velocity region is remarkably lower than $V_{GF}/M^2 = 537$ in the low velocity region, in good agree with the previous reports.[13,14] However, in all the graphene samples with bi- to octa-layered, the $V_{GF}/M^2$ ratio is reversely higher in the high velocity region as shown in Fig. 4b than that in the low velocity region as shown in Fig. 4a. This unusual trend should be addressed by the interlayer interaction in multilayer graphene.

We also compare the layer-dependence of the gas-flow-induced voltage ratio $V_{GF}/M^2$ and the measured Seebeck coefficient in Figure 4. Again, the relationship between the gas-flow-induced voltage and Seebeck coefficient in multilayer garphene becomes much more complicated than the previously believed simple proportional one. In contrast with the monotonically increasing Seebeck coefficient with layer number up to hexalayer, $V_{GF}/M^2$ rises sharply when the graphene increases from mono- to bi-layer, but drops suddenly at trilayer. Then in tetra- to hepta-layered graphene, the induced voltage $V_{GF}$ stabilized at a high level roughly around 800 $M^2$, which is comparable to the value at bilayer, but sharply drop at octalayer again, although the Seebeck coefficient at octalayer keeps the same level with hexa-and hepta-layer. The exceptional low voltage induced in tri- and octa-layered graphene should have special meaning.

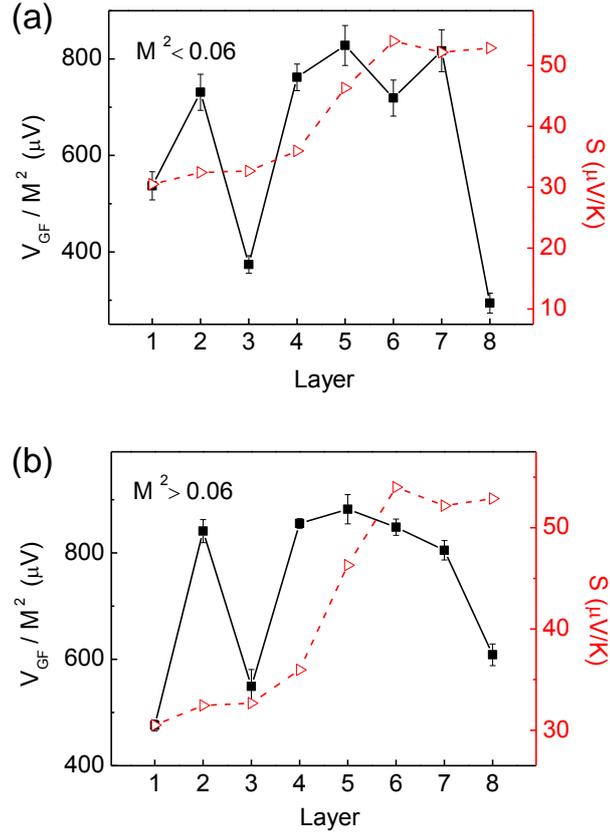

**Figure 4**. Ratio of the flow-induced voltage to $M^2$ and Seebeck coefficient. (a) $M^2 <$ 0.06, (b) $M^2 > 0.06$.

Before making further discussions on the colorful variation of gas-flow-induced voltage in the multilayered graphene, we would like to introduce our experimental exploration for thickness effect in continuous nanofilms at first. We fabricated aluminum films with thickness ranging from 10 nm to 100 nm by plasma sputtering deposition on a PET substrate. The thicknesses were detected by the quartz monitor crystal. Then we adhered copper electrodes by silver emulsion to the aluminum nanofilms to form the same sized 15 × 8 mm samples as the graphene ones. Then we conducted nitrogen gas flow experiments and the induced voltage is shown in Figure 5 against the flow velocity. The resistance and Seebeck coefficient are also measured in the same method as used for the graphene and presented in the inset of Figure 5. Although the resistance of the aluminum nanofilms decreases monotonically with increasing thickness just as in the multilayer graphene, the Seebeck coefficient (around 3.5 µV/K) as well as the flow-induced voltage are both nearly independent of thickness. This result shows from another side that the layer-dependent Seebeck effect and ability to produce voltage in gas-flow of the multilayer graphene should not be a merely thickness effect.

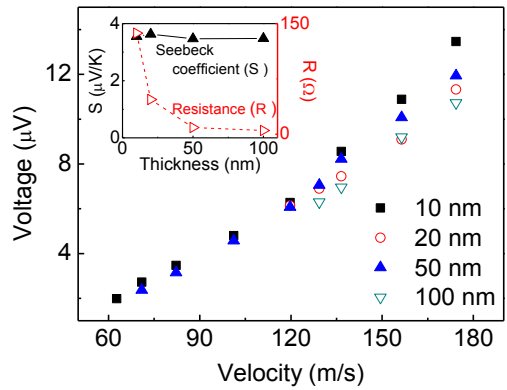

**Figure 5**. Nitrogen gas-flow-induced voltage in aluminum nanofilms with different thickness against flow velocity. The inset shows the resistance and Seebeck coefficient of the aluminum films.

It has been well demonstrated that the Seebeck coefficient or thermoelectric power of single layer graphene is dominated by diffusive carriers in the temperature range of $T < 300$ K.[27-29] The Seebeck coefficient has been demonstrated to change with the carrier density ($n$) following the semiclassical Mott relation[30] of $1/\sqrt{n}$ in a large amount of monolayer graphene samples previously.[14] However, it is clearly that the same rule is not applicable to multilayer graphene in the present work as shown above.

It have been well recognized that the CVD grown graphene should have multidomains[31,32] at the size of the samples used in the present experiment. The multilayer graphene samples through multiple transferring should have random stacking in orientation between different layers and domains. It is intuitive to reckon that the properties of the multilayer graphene should only change with layer number monotonically, but our results show strong shooting up and dropping down in both the Seebeck coefficient and flow-induced-voltage with increasing number of graphene layer. Therefore, the combining influence of graphene domains, stacking and layer number should be much more complicated and worthy of extensive theoretical and experimental investigations.

**CONCLUSIONS**

In summary, in mono- to octa-layered graphene fabricated by multiple transferring the CVD grown monolayer graphene, the Seebeck coefficient increases monotonically with the added layer and reaches its maximum value at hexa- layer, 77% higher than at monolayer，and then decrease slightly with further added layers, although the electric resistance decreases quickly to a stable value with increasing layer. In sharp contrast to the prevailing view, the gas-flow-induced voltage in multilayer graphene is not proportional to the Seebeck coefficient, but increases significantly from monolayer to bilayer, and nearly keeps in the same high level in tetra- to hepta-layered graphene, while drops down remarkably in tri- and octa-layered graphene, to value even lower than that in monolayer graphene. Therefore, the layer number, domain distribution and stacking should play joint role in the abnormal properties of multilayer graphene and worthy further exploration.

**METHODS**

In this work, single layer graphene samples were grown on 25 μm thick Cu foils (Alfa Aesar, item No.13382) by using chemical vapor deposition (CVD) method using methane as precursor.[10] The copper foil was loaded into a quartz tube, which is placed in a furnace. A typical growth process is: (1) the pressure in the growth chamber was pumped down to 1 mTorr using a vacuum pump; (2) introduce a 10 sccm flow of hydrogen gas into the chamber. (3)Then the Cu foil was heated to 1000 ℃ and annealed for 30 min to enlarge the Cu grains and remove residual Cu oxide and impurities; (4) introduce 30 sccm flow of methane gas into the chamber for 15 min with a total pressure of 56 Pa for graphene synthesis; (4) after growth the furnace was cooled down rapidly to room temperature under a 10 sccm flow of hydrogen.

The transfer process is briefly introduced as follows. Firstly, one side of the graphene/copper foil is spin coated with polymethyl methacrylate (PMMA) on at 2000 rmp for 30 s and cured at 150 ℃ for 15 min. Then the uncoated side is treated with air plasma for 5 min. After the copper was etched in FeCl$_3$ solution (1 M) for 24 h, the graphene/PMMA film was washed in HCl solution (0.6 M) for 2 h, and then in DI water for several cycles. This graphene/PMMA film was transferred onto a polyester terephthalate (PET) substrate, and dried at ambient. To improve the continuity of the graphene sample, it was redeposit with a PMMA film and cure at room temperature for 0.5 h. The PMMA was finally removed with acetone.[22]

*Acknowledgement*. This work was supported by 973 program (2012CB933403), the National NSF (10732040, 91023026, 11172124) and Jiangsu Province NSF (BK2008042) of China.

*Supporting Information Available*. Raman spectra of a monolayer graphene sample; *I-V* curves of graphene samples of different layer numbers; Flow-induced voltage in graphene samples from monolayer to octalayer at several specific velocities. This material is available free of charge *via* the Internet at http://pubs.acs.org.